# Motion of the planets: the calculation and visualization in Mathcad

**Valery Ochkov, Katarina Pisačić**
National Research University Moscow Power Engineering Institute, Russia
University North, Croatia

E-mail: ochkov@twt.mpei.ac.ru

**Abstract.** This article describes use of Mathcad mathematical package to solve problem of the motion of two, three and four material points under the influence of gravitational forces on the planar motion and in three-dimensional space. The limits of accuracy of numerical methods for solving ordinary differential equations are discussed. Usual concept of Kepler hours with uneven movement arrows illustrates Kepler's second law. Mathcad animation tools are used to illustrate solutions and links with animations are provided.

## 1. Introduction

Various software packages are built to aid physics learning and teaching [1,2] they have specific modules and solve number of problems but somewhat restricting. There are specialized physics packages such as "Motion of cosmic bodies" [3] which consists of a training simulation software and instructional materials have purpose to help students understand and explore the fundamental laws of physics and mathematical methods of investigation of the heavenly bodies movements.

In much different matter the mathematical modelling software [4,5] allow user to build mathematical model [6] of physical problem and obtain solution e.g. to solve the problem of motion of material points with a given mass (planets, satellites, comets, etc.) according to Newton's second law:

$$\sum F = m \cdot a$$

and the law of universal gravitation:

$$F = G \cdot \frac{m_1 \cdot m_2}{r_2}$$

In this paper Mathcad [7] will be used to obtain problem solutions.

## 2. Differential equation solution in Mathcad

In an environment of mathematical program Mathcad function Odesolve gives numerical solution of ordinary differential equations (ODE) [8,9], which reduces the problem of the motion of material points. If you use this feature to supplement animation tools [10], interesting and instructive solutions related to the topic of the article can be obtained.

*2.1. Example 1: Movement of three particles*

In [11] an analytical solution of three-body motion in orbit shaped as an infinity sign (or "magnificent eight" in the terminology of [3]) is given[1]. We now solve this problem numerically in Mathcad and compare the results of numerical and analytical solutions.

To numerically solve the system of ODE in Mathcad 15 it is necessary to define constants and set initial values for unknowns, initial values are set in matrix as shown in Figure 1. After defining, calculation starts with the keyword Given [8], system of equations s written using the apostrophe (keyboard shortcut Ctrl + F7) to denote the first derivative or double apostrophe for the second derivative[2], system is also including the initial conditions - position of particles and their velocity at the initial time (at the start of motion $t = 0$). In our case, scalar differential equations of motion of material points are recorded in pairs - the horizontal ($x$) and vertical ($y$) directions.

$$\begin{pmatrix} m & x_0 & y_0 & x'_0 & y'_0 \\ m & x_0 & y_0 & x'_0 & y'_0 \\ m & x_0 & y_0 & x'_0 & y'_0 \end{pmatrix} := \begin{pmatrix} 1 & 0.97000436 & -0.24308753 & \frac{0.93240737}{2} & \frac{0.86473146}{2} \\ 1 & -0.97000436 & 0.24308753 & \frac{0.93240737}{2} & \frac{0.86473146}{2} \\ 1 & 0 & 0 & -0.93240737 & -0.86473146 \end{pmatrix}$$

**Figure 1.** The initial values for the problem of the "Magnificent Eight"

In Figure 2, we can see a system of six ODE's recorded for the three material points (black, red and blue, colors are used to avoid additional notation): the product of mass and acceleration (left-hand side of the equation[3]) is equal to the sum of two gravitational forces acting on the material point of its two "neighbors". Or rather, not the actual vectors of accelerations and forces, but their projections in the horizontal ($x$) and vertical ($y$) directions[4].

Each equation can be divided by the mass corresponding to the planet, leaving only the second derivative on the left[5]. In addition, you can remove the variable *G* (gravitational constant), as in our calculations it has been adopted (for now - see the end of the article) per unit. This will simplify and speed up the calculation. But we deliberately leave these variables in the equations to maximally preserve the physics of the problem [12]: left product of the mass point and its acceleration, right forces acting on the point - the planet or satellite.

---

[1] If two points of analytical solution exist for all cases then can be reduced to an ellipse (circle), parabola and hyperbola.
[2] Here it is possible to apply the derivative operator $d/dt$
[3] Mathcad 15 has ability to give a variable a different colour and is benefited in terms instead of attributing variable indices 1, 2, 3, etc. On the other hand, the article may be published in black and white edition, where the colour variables disappear. And the colour edition can read colour-blind, that too is not very good. Once upon a time the first author sent an article about the colour in the programs in one magazine. The article was not published - the reviewer said that colour displays and printers will not soon appear and whether there will be any. And he (the reviewer) is colour-blind. Five years after the publication of this article attempts appeared language Visual Pascal, where the colour used to highlight programming constructs: built-in, user-defined, remarks, error messages, etc.
[4] Some readers, seeing the cube, not a square in the denominator of the fractions in Fig. 1 and not seeing the numerator can then recall an old anecdote: "If what fell on Newton's head is not an apple, but a more important thing (coconut, for example), the law of universal gravitation distance would be given not in the second, but the third degree."

Given

$x(0) = x_0 \quad x'(0) = x'_0 \quad m \cdot x''(t) = \dfrac{m \cdot G \cdot m \cdot (x(t) - x(t))}{\left[\sqrt{(x(t) - x(t))^2 + (y(t) - y(t))^2}\right]^3} + \dfrac{m \cdot G \cdot m \cdot (x(t) - x(t))}{\left[\sqrt{(x(t) - x(t))^2 + (y(t) - y(t))^2}\right]^3}$

$y(0) = y_0 \quad y'(0) = y'_0 \quad m \cdot y''(t) = \dfrac{m \cdot G \cdot m \cdot (y(t) - y(t))}{\left[\sqrt{(x(t) - x(t))^2 + (y(t) - y(t))^2}\right]^3} + \dfrac{m \cdot G \cdot m \cdot (y(t) - y(t))}{\left[\sqrt{(x(t) - x(t))^2 + (y(t) - y(t))^2}\right]^3}$

$x(0) = x_0 \quad x'(0) = x'_0 \quad m \cdot x''(t) = \dfrac{m \cdot G \cdot m \cdot (x(t) - x(t))}{\left[\sqrt{(x(t) - x(t))^2 + (y(t) - y(t))^2}\right]^3} + \dfrac{m \cdot G \cdot m \cdot (x(t) - x(t))}{\left[\sqrt{(x(t) - x(t))^2 + (y(t) - y(t))^2}\right]^3}$

$y(0) = y_0 \quad y'(0) = y'_0 \quad m \cdot y''(t) = \dfrac{m \cdot G \cdot m \cdot (y(t) - y(t))}{\left[\sqrt{(x(t) - x(t))^2 + (y(t) - y(t))^2}\right]^3} + \dfrac{m \cdot G \cdot m \cdot (y(t) - y(t))}{\left[\sqrt{(x(t) - x(t))^2 + (y(t) - y(t))^2}\right]^3}$

$x(0) = x_0 \quad x'(0) = x'_0 \quad m \cdot x''(t) = \dfrac{m \cdot G \cdot m \cdot (x(t) - x(t))}{\left[\sqrt{(x(t) - x(t))^2 + (y(t) - y(t))^2}\right]^3} + \dfrac{m \cdot G \cdot m \cdot (x(t) - x(t))}{\left[\sqrt{(x(t) - x(t))^2 + (y(t) - y(t))^2}\right]^3}$

$y(0) = y_0 \quad y'(0) = y'_0 \quad m \cdot y''(t) = \dfrac{m \cdot G \cdot m \cdot (y(t) - y(t))}{\left[\sqrt{(x(t) - x(t))^2 + (y(t) - y(t))^2}\right]^3} + \dfrac{m \cdot G \cdot m \cdot (y(t) - y(t))}{\left[\sqrt{(x(t) - x(t))^2 + (y(t) - y(t))^2}\right]^3}$

**Figure 2.** System of ODE's written in Mathcad

In Mathcad 15 ODE block function ends by writing Odesolve function (Fig. 3a) which has syntax Odesolve[6] ([vector], x, b, [intvls]) following arguments [12]:
- [vector] is optional, used only in system of ODEs and represents vector of function names, as they appear within the solve block is a solution of ODE (within Mathcad Prime it's necessary for the function name in the brackets indicate its argument - see below).
- x is variable, which is conducted by numerical integration (the argument of the unknown functions, within Mathcad Prime this variable is not specified);
- b is "right" terminal point of integration interval ("left" value is zero)
- [intvls] is optional argument and represents the number of points on which is tabulation of the unknown functions conducted (defaults is 1,000 points).

---

[5] It's not worth doing because in this case there will be an interesting effect in the numerical solution of the problem. The problem of the three points can be reduced to the problem of the two points, giving the mass of a planet zero and giving it some initial velocity. Two planets will move in their orbits, and the third (with zero mass) in a straight line. But if in the equations of mass cut, the straight line somehow become curve
[6] In Mathcad there are other tools for numerical solutions of differential equations (ordinary and partial), but in this article are not considered.

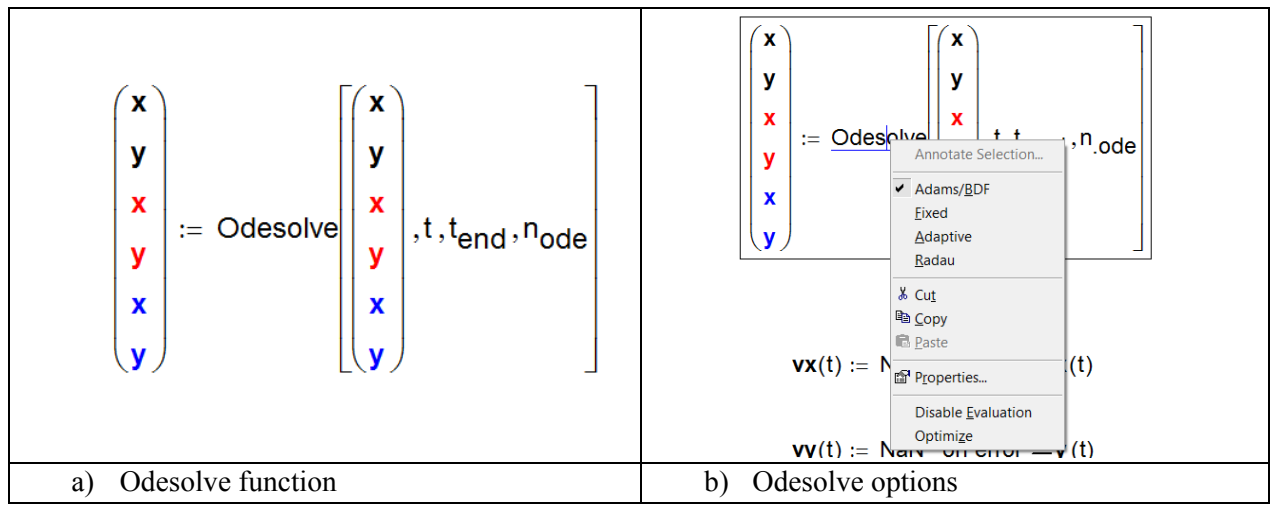

| a) Odesolve function | b) Odesolve options |

**Figure 3.** Writing and setting Odesolve

In addition, by pressing the right mouse button over function Odesolve (Fig. 3b) you can select one of four possible methods for the numerical solution of the problem:
- Adams / BDF - hashing algorithm Adams and backward differentiation formulas,
- Fixed (fixed-step integration, you can see a comparison of this method with the Euler method [H]),
- Adaptive (integration with variable pitch [13]),
- Radau algorithm for stiff ODE.

If three material points set initial conditions, as shown in Figur e 1 taken from [11], the solution of the ODE system (Fig. 2) will be a function whose step-to-step graphical representation is shown in Figure 4 a).

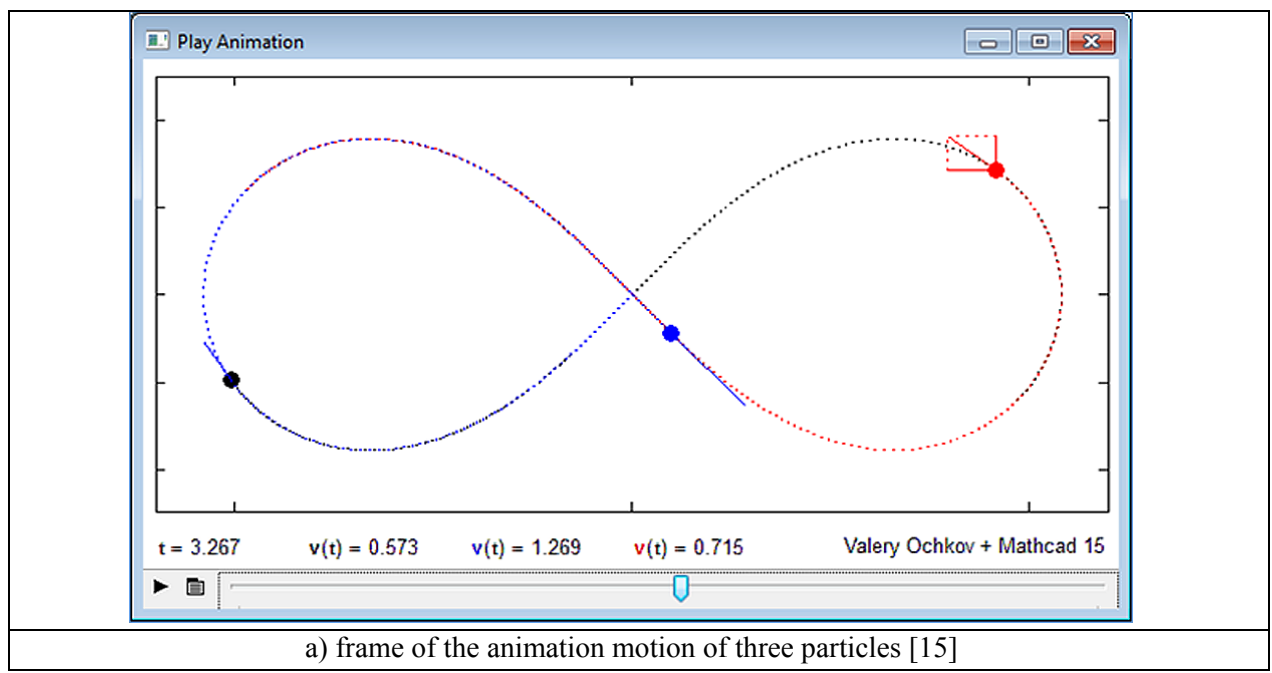

| a) frame of the animation motion of three particles [15] |

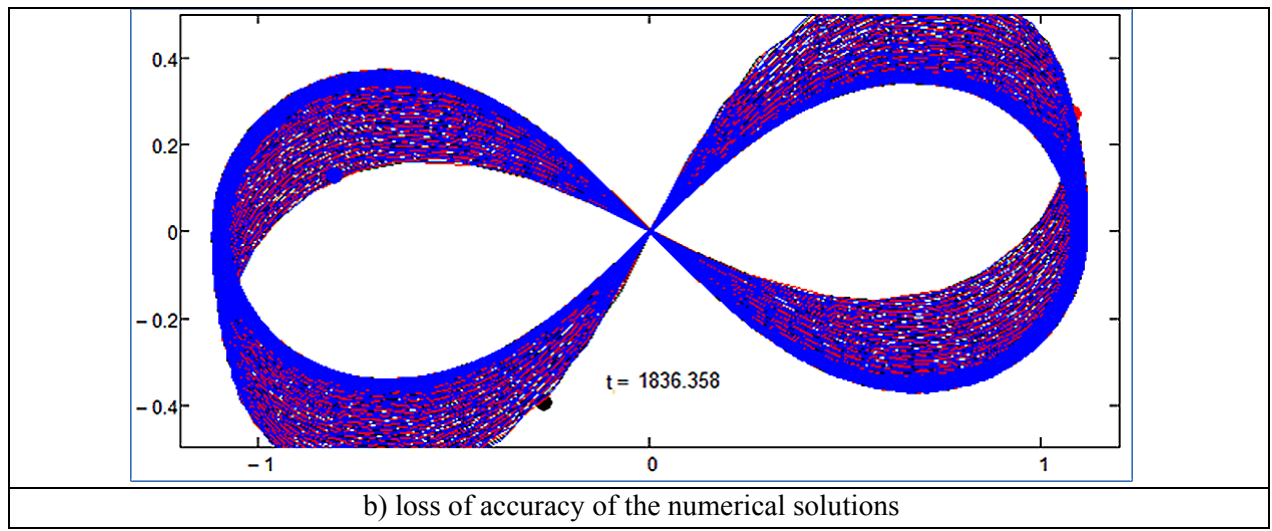

b) loss of accuracy of the numerical solutions

**Figure 4.** Three particles in orbit shaped as infinity sign

Comparison of "telemetry data taken from orbit" shown in Fig. 4a, with the data of [11] gives a complete coincidence of the results, which indicates the adequacy of our model, recorded in the equations of Fig. 1. But to be precise, the following happens: at first, about a hundred "turns" there is no deviation noticed, but with increasing values of $t$ as a result of the accumulated error of the numerical solution of the problem there is a certain "hollowing out" of the orbit (Fig. 4b). This feature is clearly seen in the second example.

*2.2. Example 2: Movement of four particles*

In the corners of a square with sides equal to one, placed four material points (planets) with the same mass. The initial velocities of the points are equal in absolute value and have the following directions: blue planet – up, red – down, black - to the right, while green - to the left. Motion of the planets with the starting position (in [3] this motion is called a round dance) shown in Fig. 5.

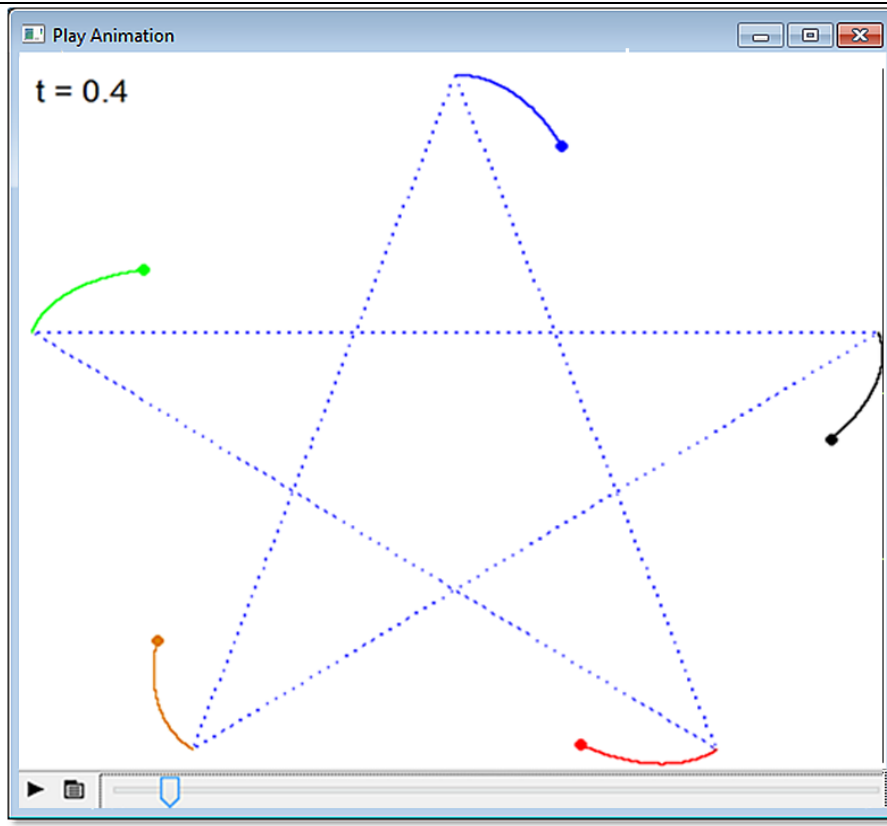

a) Motion start points

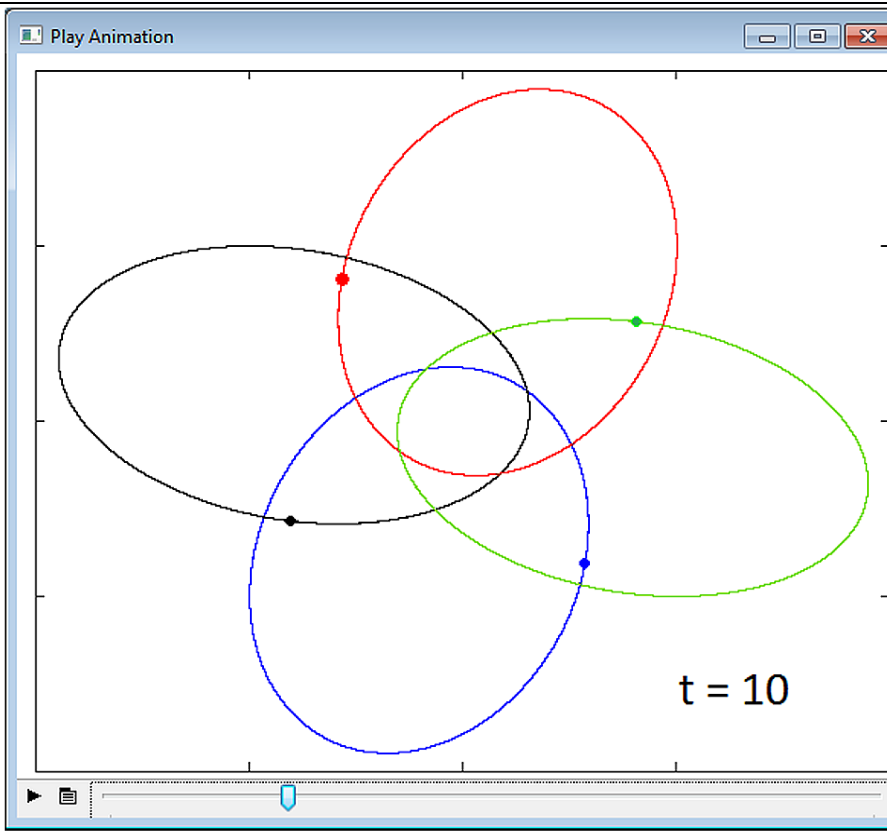

b) Full match with the analytical solution

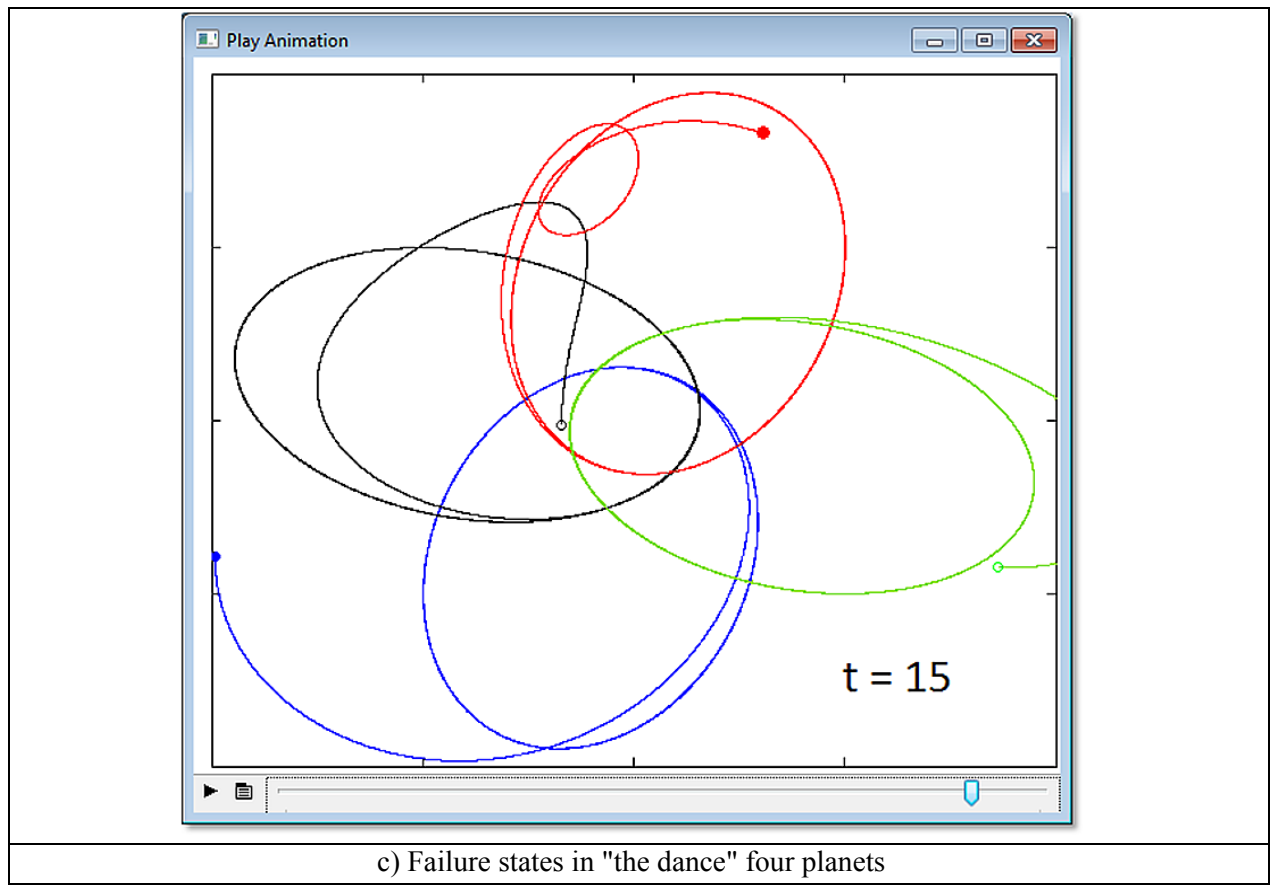

c) Failure states in "the dance" four planets

**Figure 5.** Frame animation dance of four planets [16]

"Cosmic Dance", like any other dance cannot last indefinitely - at some point, dance crumbles, and its members scatter in different directions. Something similar can be seen in the dances of our four planets with increasing value of the variable numerical integration *t* - see Fig. 5a.

The motion of material points can be made in the third dimension, moving out of the plane in the volume ("space"[7]). To do this it is necessary and sufficient to write another three equations in z direction in the system shown in Figure 1, add the initial data and solution vector will have z component for each particle. Figure 6 shows the "dance" of three planets in volume. The third axis is directed upwards (it is given the same initial uniform motion of the planets up along the axis z), and the curves are graphs of the three solutions[8]. This graph can also be considered as a change in the location of points on a plane (see. Fig. 4a) over time (such graphs are called integral curves).

---

[7] We say "flying in interplanetary space." A more correct to say you need to "fly on the plane interplanetary." Most of the flight calculations of natural and artificial space objects is carried out in two dimensions by choosing the right plane.
[8] These drawings often depict young children: first, everything is going well, and then the child gets tired to draw properly and draws on nearly finished picture "Kalyaki-Malaki" a Russian name for children scribble.

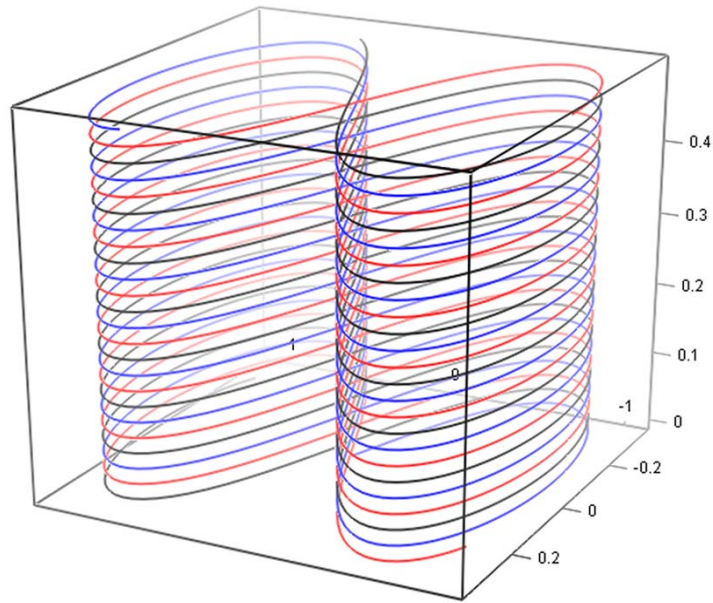

**Figure 6.** A three-dimensional problem of the motion of three particles [17]

Figure 4a shows one of the three projections of the motion of three points (Fig. 6) when the *z* axis is directed toward the observer. Trajectory displayed in Fig. 6 can be interpreted as a wrap of three strands of hair in a pigtail, which is clearly seen in animations of these trajectories. If the initial velocity for the three planets was different, result would be confusing path of movement on which would be difficult to exclude the component associated with the loss of accuracy of the numerical solution of the problem.

But back to the "flat" cases, bearing in mind, firstly that the planets of the solar system move in orbits located on the same plane, and, secondly that the numerical methods built into Mathcad, with God's help cope with the plane problems.

Two of our previous examples ("magnificent eight" and "cosmic dance"), as repeated tests verify the coincidence of numerical and analytical solutions, while the error of numerical methods have not yet reached a critical level. Adequate[9] solution for not too large values of *t* allows us to apply this method (see Figures 1 and 3) to investigate the case that yet has no solution, for which analytical solutions are unknown and impossible. To do this it is necessary and sufficient to add new equations with the new expression on the right, to change the initial conditions, and select the appropriate method for the numerical solution.

*2.3. Example 3: Intercept satellite*

The red planet with blue satellite is moving toward the black planet. What happens to these celestial bodies at the moment of convergence? To investigate this case we have a few numerical experiments. We solve this problem using two different numerical methods. Fig. 7 displays the frame of the animation, which shows that at the time of convergence of the planets moon changes it's "patron" and begins to rotate around the arc (black) of the planet. The red planet is continuing its celestial path alone.

---

[9] What is numerically "not too large value".

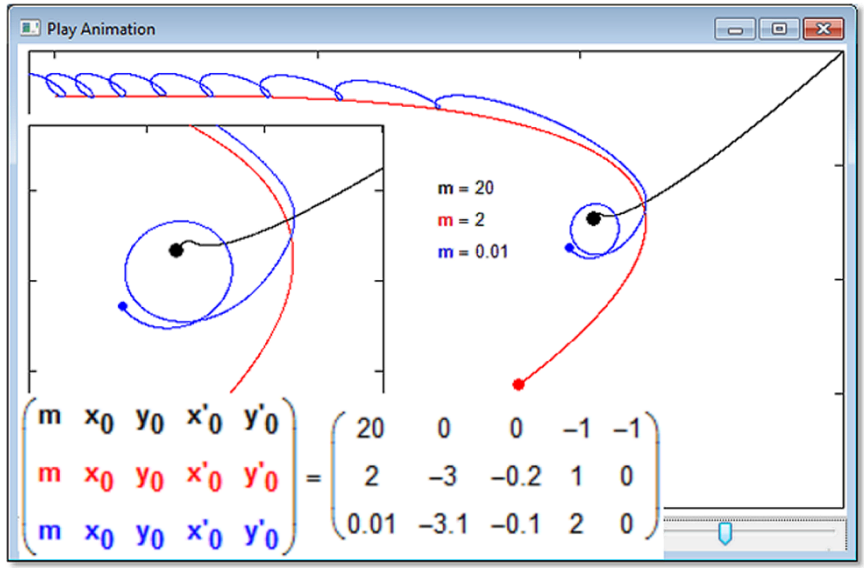

**Figure 7.** Frame animation of satellite interception [18]

When creating animation of satellite interception, shown in Fig. 7, was model was build using the Radau method (see Fig. 3) within Odesolve function. If the decided to change the method from Radau to Adams / BDF, a completely different picture will be seen: the interception of satellite does not take place, as will be observed only some correction of its orbit due to the proximity of the black planet to the "red planet - blue moon" - see Fig. 8.

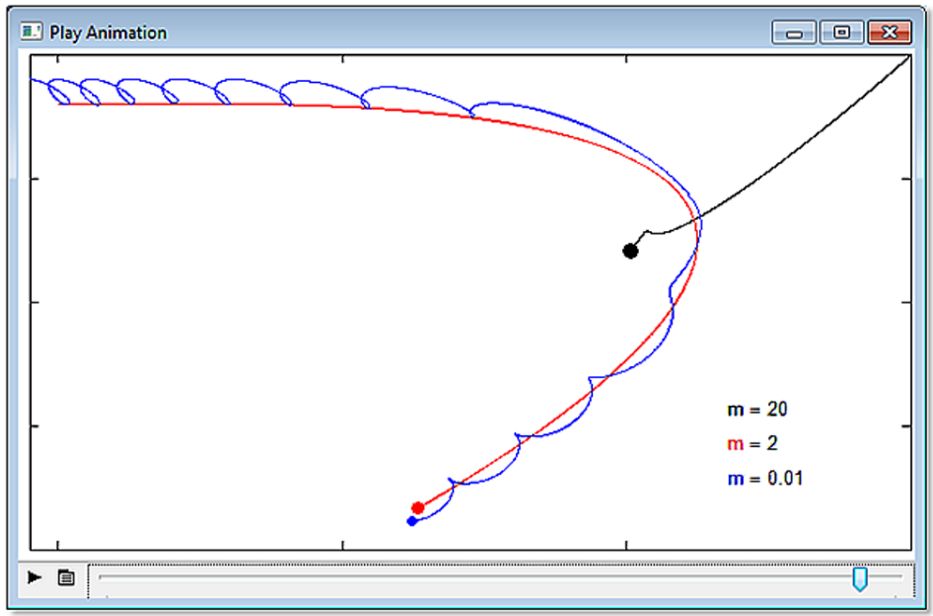

**Figure 8.** Correction of satellite orbit [18]

Here the author, honestly speaking, wanted to quit writing article about the calculations as of interesting and important tasks, when the results are highly dependent on the solution methods and fundamentally changing with the method and not only quantitatively (see Fig. 4 and 5) but also qualitatively. See [19] and the end of the article.

Animations displayed in one frame in Figures 5, 6, 7 and 8, again focus our attention on the "glitter and poverty" of numerical methods for solving ODE: they allow you to somehow solve the problems, which are "too tough" to solve using analytical methods, and create quite entertaining and instructive animation - such, for example, which is shown in Fig. 7, 8 and 9 below but they can "lie us straight in the eye." Still, we'll get to the bottom of this problem.

*2.4. Example 4: Exchange of satellites*

Figure 9 shows another non-trivial and more complex case - the exchange of satellites. Cases previously shown in Figures 7 and 8 are everyday analogy: there is a certain married couple (the system "planet-satellite"), which is approaching a foreign object, or rather, the subject. He can either "discourage" the satellite (Fig. 7), or add to the life of this couple some "perturbations" (Fig. 8). When published online animation shown in Fig. 9 [20], it was commented: "In life, too, is common: two couples meet, come together and eventually realize they have made the wrong choice, take the law (two divorces and two new wedding) or informal Swing." This comment was immediately responded by site moderator asking user to remove, as "Minors also visit this site." Nevertheless here is shown such cosmic motion.

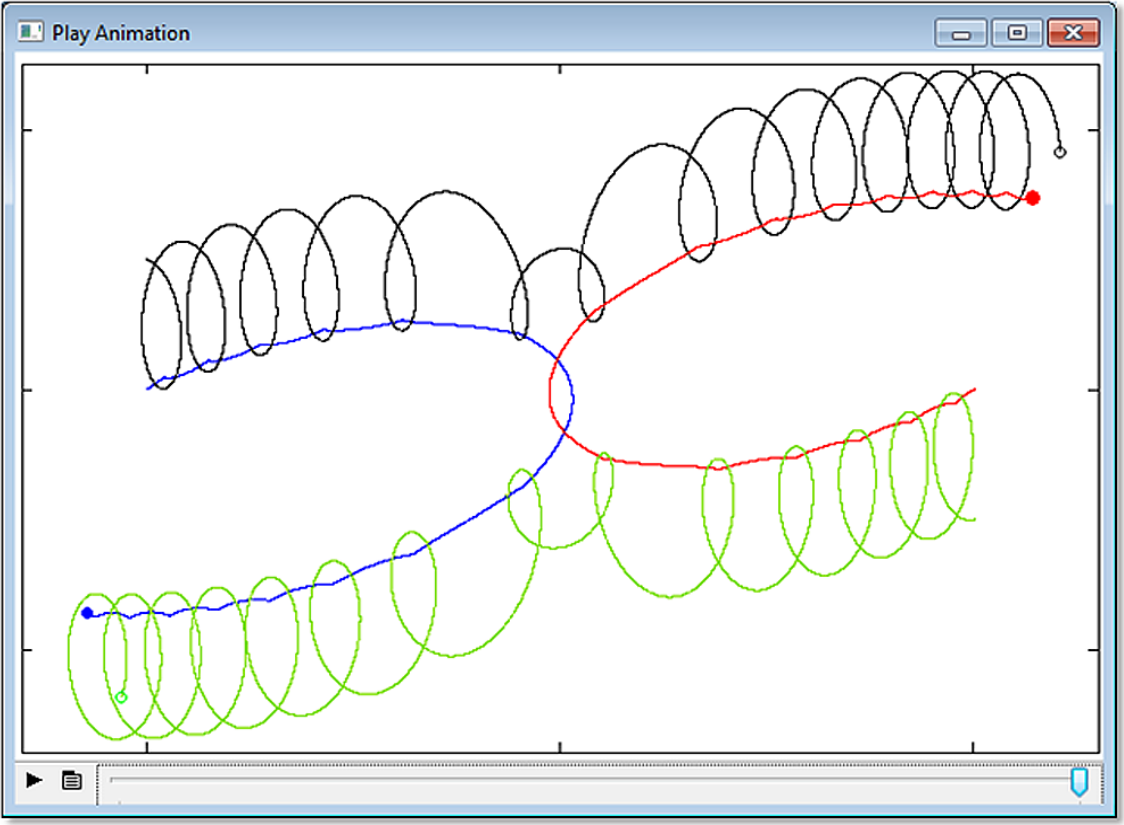

**Figure 9.** Frame animation of satellite's exchange [20]

In Figure 7, 8 and 9 is shown somewhat frivolous behavior and cannot be taken seriously about "everyday" trajectories due to uncertainty of the numerical methods used to solve these problems. But it is interesting, and based on the "adequacy" of physics. And with that, we have to agree.

*2.5. Example 5: Kepler's watch*

The new version of Mathcad - Mathcad Prime does not yet have animation, but can solve the ODE with the units of measurement of physical quantities. Prior to that, these units were ignored, which is prone to errors: if somewhere in the system of equations power of three was replaced by power of two, the system cannot be solved or resolved, but gives the wrong answer.

Figure 10 shows the rotation of the satellite orbit with the real values of distance, speed and most importantly, the gravitational constant G. Pick up time of such initial data to the rotation period of the satellite was equal to 12 hours. In addition, the graph added sector elliptical orbit that illustrate Kepler's second law: the area of these sectors in the regular (hourly) intervals are equal (to within hundredths of a percent - see. Fig. 10b), and the numerical solution of the problem, which is an additional proof of loyalty our model[10]. The very same animation can be seen as yet another unusual mathematical watch [21], in which the arrows move in elliptical dial with variable speed[11]. These watches will look good in the museum Kepler, planetariums, observatories etc.

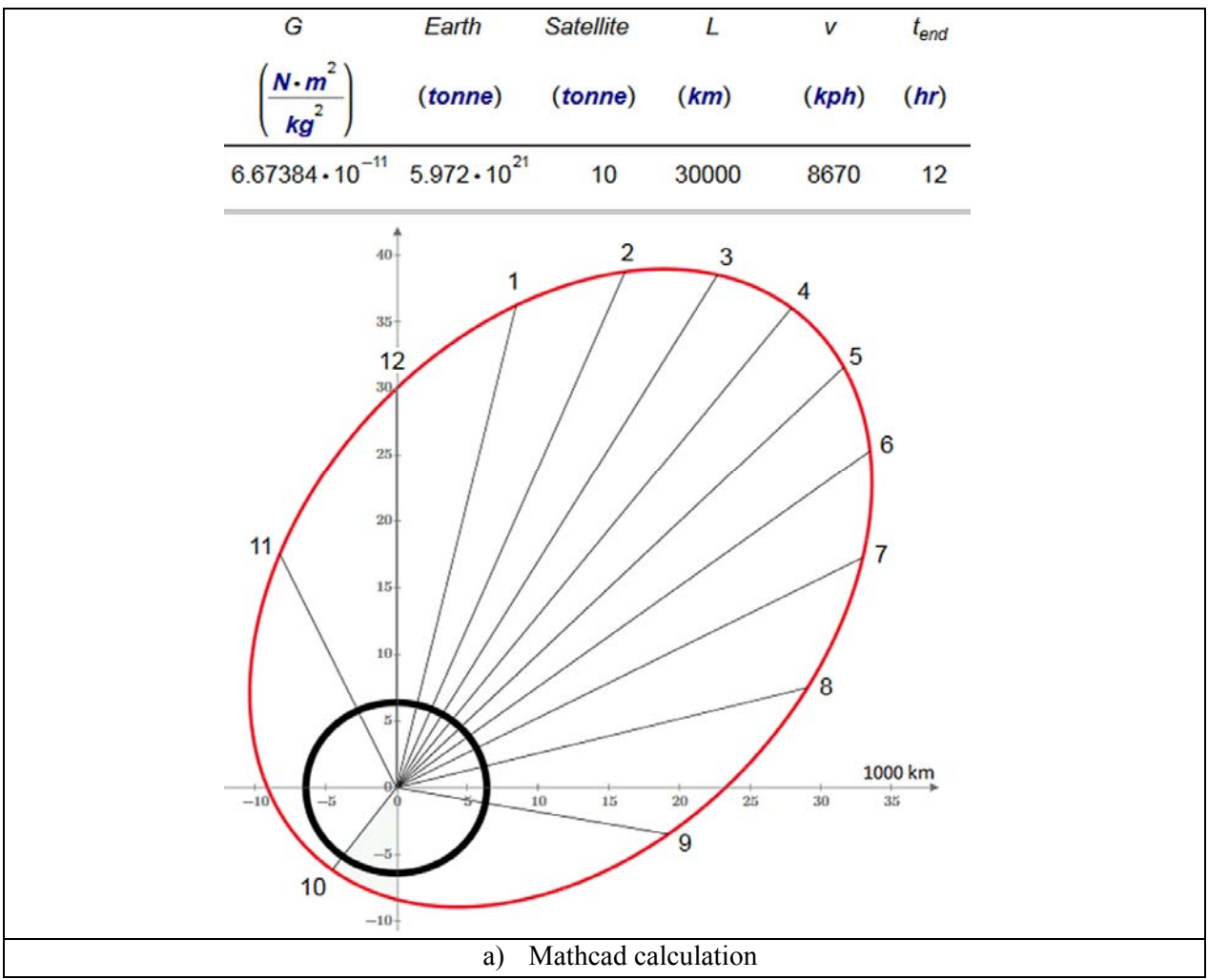

a) Mathcad calculation

---

[10] The problem of two bodies has an analytical solution, in particular, an ellipse as trajectory. Our ellipse can be reduced to a canonical form and calculate the values of its parameters - the coefficients of second-order curve, axes, and other tricks.

[11] Try to make such invention. Kepler watch is traditional clock with hands and dial. The difference lies construction of a clock face, Kepler's is not a circle but an ellipse with the numbers 1, 2, 3 ... 12 (I, II, III ... XII). Arrows (hour, minute and second) move in a circle uneven. The length of the arrows also changes as they move. Arrowheads describe elliptical trajectory. Additionally the dial displays sector, covering two adjacent hours between which at any given time is second hand. This sector is changing its location and shape of every five seconds, but its area remains constant according to Kepler's second law. Around the site of attachment of arrows, traced a circle representing the Earth, and at the end of the second hand is placed a circle showing satellite. These watches can hang in mathematics classrooms, planetariums, observatories, museums, attracting people's attention, causing them to think about the laws of celestial mechanics, in particular, the second law of Kepler.

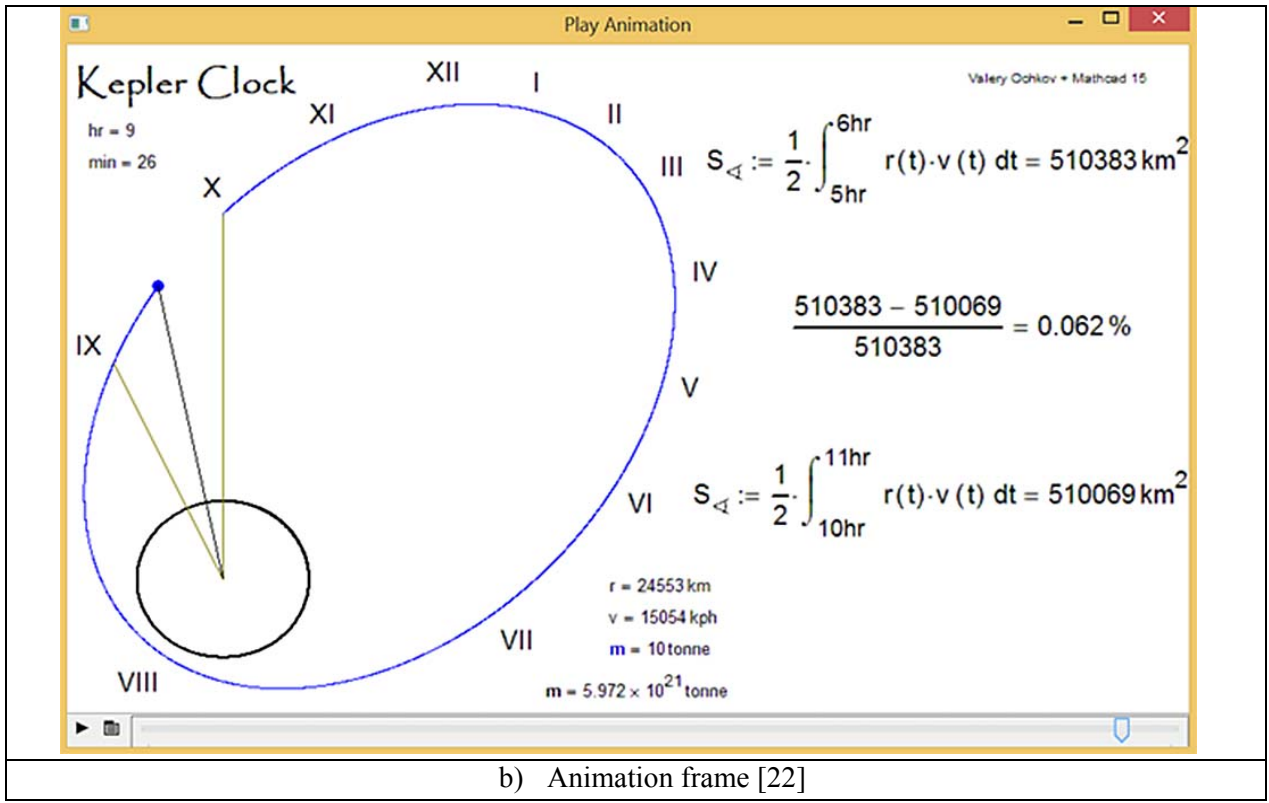

b) Animation frame [22]

**Figure 10.** Kepler's Watch

Figure 10 shows that the satellite at "eight hours" is as close to Earth (perigee) and at this point should have according to the second law of Kepler's maximum speed. In the same area two hours (apogee) the speed of the satellite is at a minimum.

But near the Earth can depart from the law of universal gravitation; move on to the potential field of gravity close to our planet, characterized by the acceleration of gravity, and take account of its atmosphere. Such a task is also easy to solve with the assistance of Mathcad - see. Fig. 11.

*2.6. Example 6: Parachute jump*

Formulation of problem of parachute jump arose under the influence of the famous parachute jump from a height of about 40 km (from the stratosphere) Austrian Felix Baumgartner [23]. Such jump from even greater height in October 2014 repeated Alan Eustace - CEO of Google [24].

| $h_1$ | $r_1$ | Mass | $h_2$ | $r_2$ | k | $t_{end}$ |
| (km) | (cm) | (kg) | (m) | (m) | | (min) |
| --- | --- | --- | --- | --- | --- | --- |
| 30 | 30 | 110 | 1500 | 2.5 | 1.7 | 15 |

Changing the air density at altitude

$\rho_{Air}(h_1) = 0.01002 \, \frac{kg}{m^3}$

$\rho_{Air}(h_2) = 1.058 \, \frac{kg}{m^3}$

$\rho_{Air}(0 \, m) = 1.225 \, \frac{kg}{m^3}$

$$\rho_{Air}(h) := \begin{Vmatrix} L \leftarrow 0.0065 \, \frac{K}{m} \\ T_0 \leftarrow 288.15 \, K \\ T \leftarrow T_0 - L \cdot h \\ M \leftarrow 28.9644 \, \frac{gm}{mole} \\ p \leftarrow 101325 \, Pa \cdot \left(1 - \frac{L \cdot h}{T_0}\right)^{\frac{g \cdot M}{R \cdot L}} \\ \frac{p \cdot M}{R \cdot T} \end{Vmatrix}$$

Changing the height range, cross-sectional area and volume of the skydiver

$r(h) := if(h > h_2, r_1, r_2)$  $Section(h) := \pi \cdot r(h)^2$  $Volume(h) := \frac{4}{3} \pi \cdot r(h)^3$

Solve

$$h(0 \, c) = h_1 \quad v(0 \, c) = 0 \, \frac{m}{s}$$

$$v(t) = h'(t)$$

$$Mass \cdot v'(t) = k \cdot \rho_{Air}(h(t)) \cdot Section(h(t)) \cdot v(t)^2 + g \cdot Volume(h(t)) \cdot \rho_{Air}(h(t)) - g \cdot Mass$$

$$\begin{bmatrix} h \\ v \end{bmatrix} := odesolve\left(\begin{bmatrix} h(t) \\ v(t) \end{bmatrix}, t_{end}\right)$$

$t := 0 \, s, \frac{t_{end}}{1000} \ldots t_{end}$

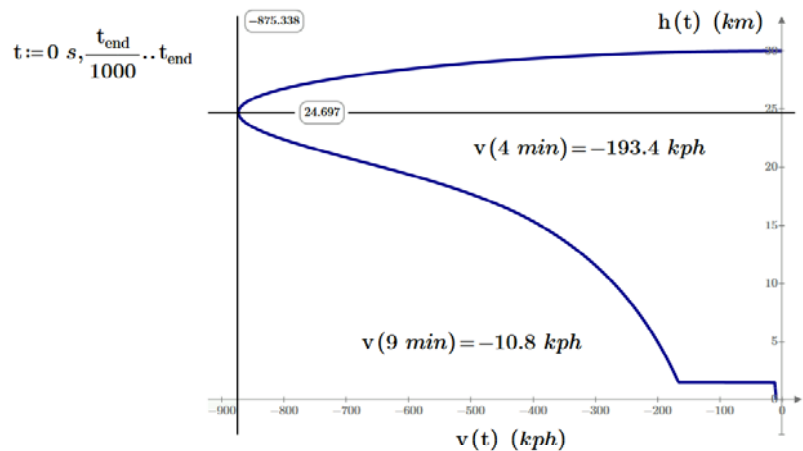

$v(4 \, min) = -193.4 \, kph$

$v(9 \, min) = -10.8 \, kph$

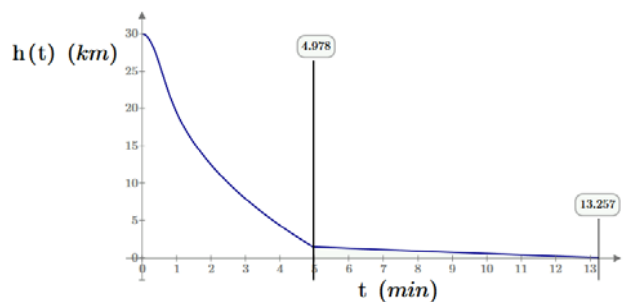

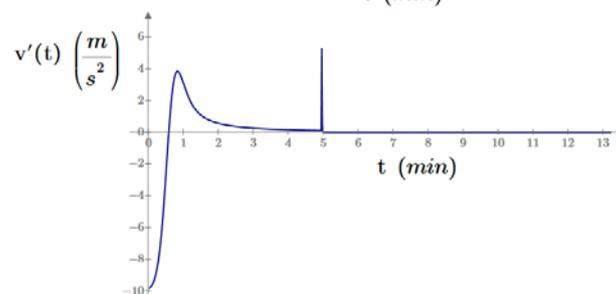

**Figure 11.** Parachute jump mathematical model

Consider a mathematical model: a parachutist at the beginning of the flight - a sphere of radius $r_1$ and a mass of 110 kg. After opening the parachute, he turns into a ball of radius $r_2$. The mass of a skydiver is not changed. Parachutist jumps from a height $h_1$, and opens the parachute at a height $h_2$. These raw data are entered into Mathcad calculation in the form of a table (Fig. 11), in which the first line are the names of the variables, the second - the unit of measurement, while the third - the numerical values.

The task included additional background information: the air density as a $\rho_{Air}$ function of height, the coefficient of friction of the air against parachutist *k*, and the estimated time of the flight parachutist tend. After the input data three auxiliary functions were recorded: change of ball parachute radius *r* depending on the altitude range, the cross-sectional area "*Section*" and its volume "*Volume*". These functions have the form of steps: opening the parachute to ($h > h_2$) they return one value, and after expansion - other.

In Figure 11 it can be seen differential equation, "equalling" the forces acting on the parachutist. This equation is nothing but a mathematical expression of Newton's second law, which states that the sum of the forces acting on a body is equal to the product of its mass and acceleration.

What forces are acting on the parachutist? The first force - its weight: the product of mass and acceleration of free fall g. This force is directed downward, which is why it is in the equation with a negative sign. The second force - a force of air resistance, which is usually neglected when considering the flight (fall) of stone, but which cannot be neglected in the case of a paratrooper, otherwise it will break into a pancake. We assume that this force is proportional to the density of air, cross-sectional area of a falling body and the square of its velocity. The coefficient of proportionality is a variable *k*. The third force - Archimedes force - the weight of the air parachutist displaced. If the parachute has not yet been opened, this force can be ignored, but if our paratrooper opened the parachute, "bloated" (according to our model) to a ball with a five-meter diameter and weighing a hundred pounds, then you need to take into account this effect. These two forces are directed upwards, so in their formulas, we see a plus sign. All three of these forces, we repeat, are balanced by the force of inertia - the product of mass and acceleration.

The problem can complicate - take into account, for example, changing the value of acceleration of gravity height, not spherical, but a more complex form of flying parachute, changing the value of k depending on the mode flow around the body (laminar or turbulent - all the subject matter of science of aerodynamics), the horizontal component of the parachute flight, due to the speed of the aircraft from which it sprang, and/or wind speed. But even without this our calculation turned out quite plausible: skydiver in free flight for about five minutes gradually picks up speed to 875 km/h, and then the speed drops due to increased air density. After opening the parachute paratrooper speed decreases sharply to 11 km/h. At this rate, he landed after 13 minutes flight. You can remove one more assumption. At $h > h_2$ parachute opens instantly. In reality, this is for some fraction of a second that can be accounted for by making the function *r(h)* more complex: not a step, but a kind of slope.

*2.7. Example 7: Jump from the tower into the water*

Figure 11 shows the model (and its implementation in the Mathcad environment) of parachute jump to Earth, rather regarded as jump to the ground. But parachute could not only land, but also splash. Figure 12 shows the calculation of the parameters of a jump from a tower. There is also a function-step, not related to the volume of "jumper", but to the density of the medium: air or water. In the calculation, there is another step function - built in Mathcad function named **sign**, which returns the sign of the argument: minus one for negative, zero for zero and plus for the one with a positive argument. This built-in feature allows us to take into account that the effect of resistance of the medium always acts in the opposite direction of the velocity. The sign of velocity values is lost due to the raise of this quantity to square.

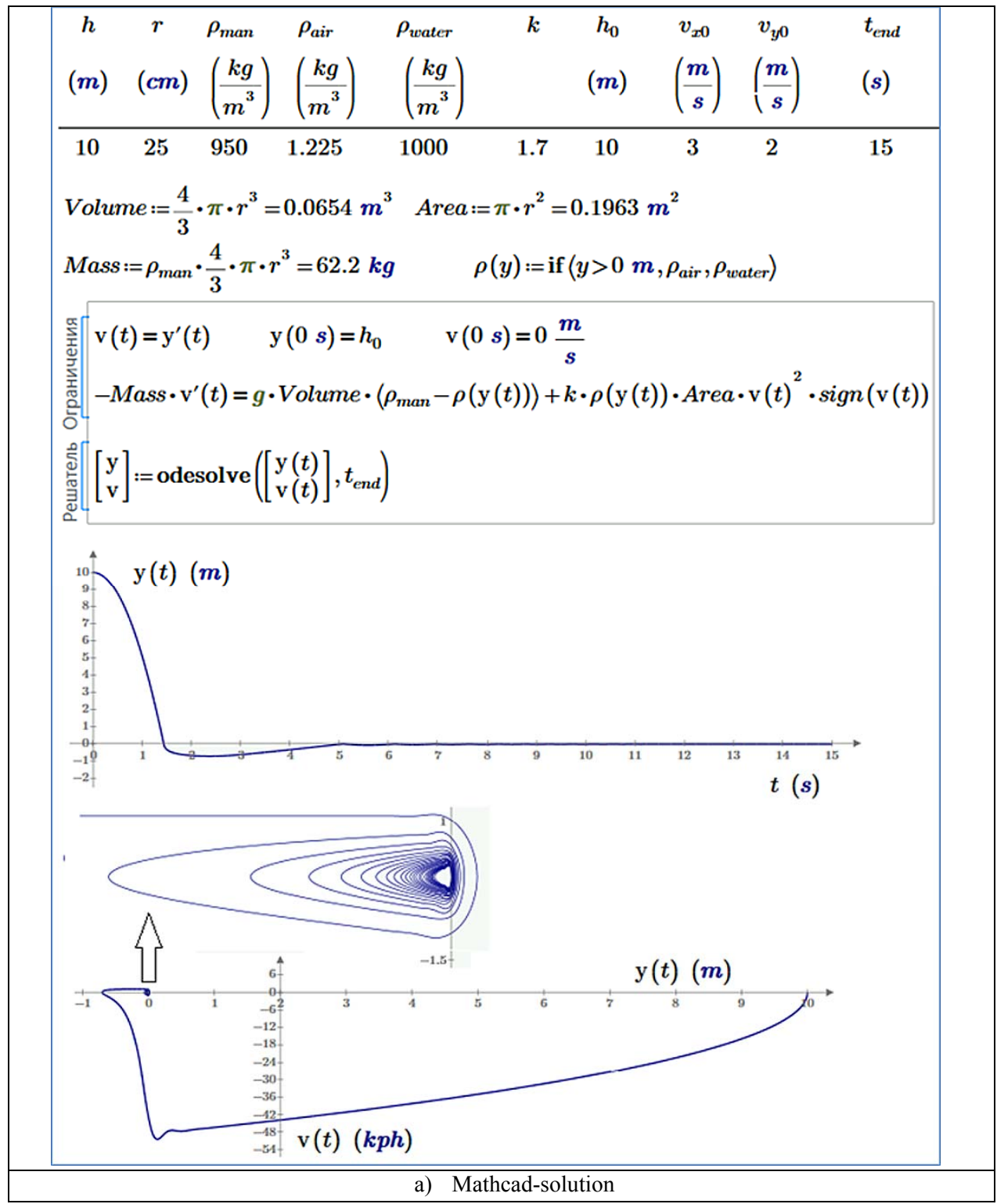

a) Mathcad-solution

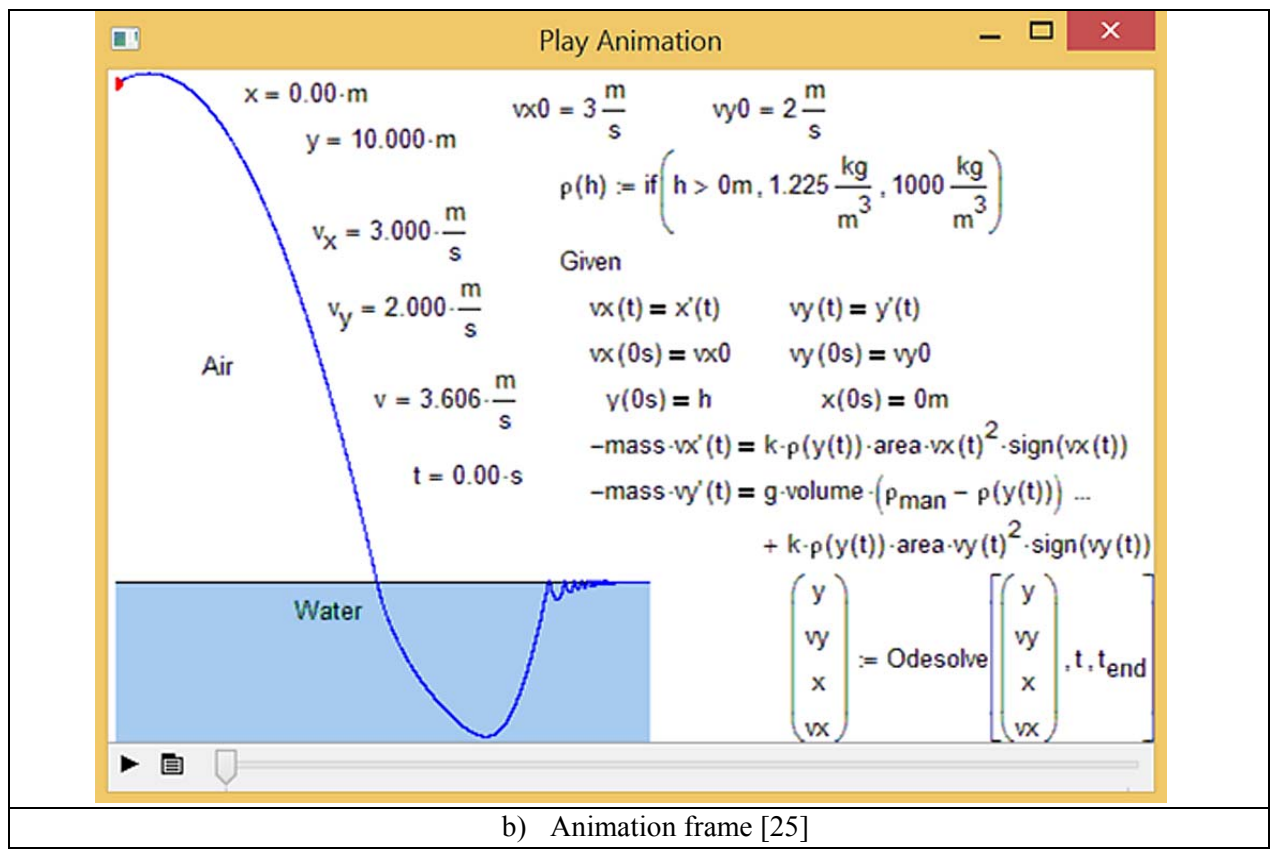

| b) Animation frame [25] |
|---|

**Figure 12.** Simulation of the jump from the tower into the water

In figure 12a parameters of jump from the tower into the water is shown. One of the graphs displays so-called attractor: jumper emerging from the water will perform damped oscillatory motion on its surface. Animation of this model (Fig. 12b) located at [25].

However, in the water cannot only dive - from the water we also can "jump." Going back to the main topic of the article, there will be another appropriate "missile" example.

*2.8. Example 8: Launch of the submarine.*

In Figure 13 we can see an animation frame of vertical take-off and the subsequent fall in the water after missile launched from under the water. This mathematical model differs from previous ones in the way of forces acting on the material point, adding thrust rocket engine, as well as the fact that the mass of a material point is reduced due to fuel and oxygen consumption.

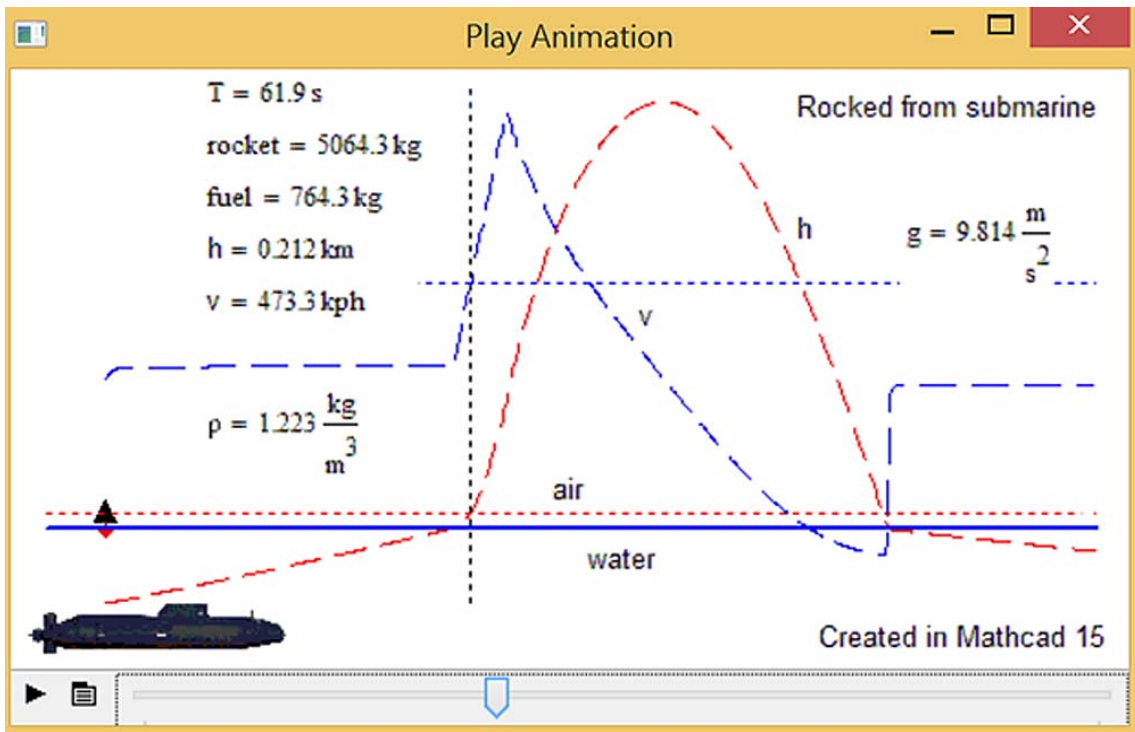

**Figure 13.** Animation Frame of the missile from a submarine

Figure 13 also shows the graphs of the height the rocket (h - red dotted line) and speed (v - blue dotted line ) depending on the time of flight.

**3. Conclusion**

The above tools allow us to easily, quickly and with sufficient accuracy, model and animate the various cases of motion of material points and bodies under the action of gravity, inertia, engine thrust and environmental resistance. Readers are invited to consider the model: spacecraft in Earth orbit includes rocket engine with known traction, starts to the moon and after take-off with the engine off goes into orbit of the Earth satellite.

If the engine accelerates against the speed of the satellite, it "will enter the atmosphere", either burn or successfully land or fly down the help of a parachute. Elements of these cases, the motion of bodies we have considered above.

In this case, you need only to remember one important thing. To paraphrase a famous saying, we can say that there is "a lie, a blatant lie ... and numerical solution of differential equations." This can be confirmed once again, looking at Figures 5 and 6, 7 and 8. However, numerical methods for solving differential equations have long been used for space missions, wherein the telemetry data is used to correct the motion which was calculated numerically in advance with certain accuracy. So we are taking to execution, someone's request or order, understand that the request or orders can be intentional or not the lie that we are in the process of levelling the execution of a request for additional information.

At present, many physical practicums at schools and universities are equipped with computers with multimedia projectors. In this practicum can be conducted physical experiments (to study, for example, the swing of the pendulum), showing on the big screen as an experiment, so that everyone can see everything well. On the same screen, you can show the solution of the differential equation of oscillations of a pendulum, to compare the real physical phenomenon with its mathematical model, explaining the differences and limitations of simplified models.

Equations - algebraic and differential – are terrible for students not in themselves but their solution methods. Now the computer can be quite simple used to solve such equations. The main task is to make the equation or system of equations, understanding the physics of the problem. Therefore, these tasks cannot be "torture" but the pleasure for students and teachers.

On the site [26] authors have placed a large number of different dynamic problems with their solutions in Mathcad and animation tools Mathcad [27]. Here are some of them:
- single swing of the pendulum;
- coupled pendulums swing;
- rotation of planets with satellites (the theme of this article, these decisions subject of a separate subforum [28];
- the launch of the submarine;
- human diving into the water;
- toboggan slide down the hills;
- movement of the car;
- movement through an underground tunnel direct gravitational train et al.

Such resources can be used as prefinished materials for teachers not only to teach advanced features of Mathcad but also to demystify physical phenomenon as well as mathematical models and solution methods.

Short post scriptum: Dream of the first author was to become astronaut, but due to imperfect sight the dream never came true. Running such calculations are way to come closer to this dream.